\documentclass[aps,epsfigure,twocolumn,superscriptaddress,showkeys,nofootinbib,floatfix]{revtex4-1}
\usepackage[colorlinks=true,linkcolor=blue,urlcolor=blue,citecolor=blue,pdfusetitle]{hyperref}
\usepackage[utf8]{inputenc}
\usepackage[english]{babel}
\usepackage{amsmath}
\usepackage[caption = false]{subfig}
\usepackage{braket}
\usepackage{graphicx,epstopdf}
\usepackage{blindtext}
\usepackage[table,xcdraw]{xcolor}
\usepackage{lipsum}
\usepackage{amsfonts}
\usepackage{bbm}
\usepackage{amssymb}
\usepackage{enumerate}
\usepackage{color}
\usepackage{latexsym}
\usepackage{times,txfonts}
\usepackage[normalem]{ulem}
\graphicspath{ {./figures/} }
\usepackage{graphicx}

\usepackage{amsmath}

\begin{document}

\title{Quantum circuit evolutionary framework applied on set partitioning problem}

\author{Bruno O. Fernandez}
\email{brunoozielf@gmail.com}
\affiliation{QuIIN - Quantum Industrial Innovation, EMBRAPII CIMATEC Competence Center in Quantum Technologies, SENAI CIMATEC, Av. Orlando Gomes, Salvador, BA 1845, Brazil.}
\affiliation{Latin America Quantum Computing Center, SENAI CIMATEC, Av. Orlando Gomes, 1845, Salvador, BA, Brazil, CEP 41850-010.}

\author{Rodrigo Bloot}
\email{rgbloot@gmail.com}
\affiliation{Federal University of Latin-American Integration, Foz do Iguaçu, PR, Brazil.}

\author{Marcelo A. Moret}
\email{moret@fieb.org.br}
\affiliation{QuIIN - Quantum Industrial Innovation, EMBRAPII CIMATEC Competence Center in Quantum Technologies, SENAI CIMATEC, Av. Orlando Gomes, Salvador, BA 1845, Brazil.}



\begin{abstract}
Quantum algorithms are of great interest for their possible use in optimization problems. In particular, variational algorithms that use classical counterparts to optimize parameters hold promise for use in currently existing devices. However, convergence stagnation phenomena pose a challenge for such algorithms. Seeking to avoid such difficulties, we present a framework based on circuits with variable topology with two approaches, one based on ansatz-free evolutionary method known from literature and the other using an introduction of an ansatz with circuital structure inspired by the physics of the Hamiltonian related to the problem, considering a, named here, pseudo-counterdiabatic evolutionary term. The efficiency of the proposed framework was tested on several instances of the set partitioning problem. The two approaches were compared with the Variational Quantum Eigensolver in noisy and non-noisy scenarios. The results demonstrated that optimization using circuits with variable topology presented very encouraging results. Notably, the strategy employing a pseudo-counterdiabatic evolutionary term exhibited remarkable performance, avoiding convergence stagnation in most instances considered. This framework circumvents the need for classical optimizers, and, as a consequence, this procedure based on circuits with variable topology indicates an interesting path in the search for algorithms to solve integer optimization problems targeting efficient applications in larger-scale scenarios.
\end{abstract}

\maketitle

\section{Introduction}
Quantum computers hold promise for solving problems that are intractable on classical computers. Pioneering work \cite{Deut,bernstein1993quantum,shor1994algorithms,grover1996fast} boosted deeper studies on the construction of quantum devices. In this direction, optimization problems are of great interest for the possible use and utility of Noise-Intermediate-Scale-Quantum (NISQ) devices (see, e.g., \cite{preskill2018quantum,Abbas}).
Several authors have studied techniques and algorithms to be used in the NISQ era. In particular, much attention has been given to variational methods in recent years \cite{vqa}. Such methods are based, in general, on fixed circuits with varying parameters. The Variational Quantum Eigensolver (VQE) \cite{vqe_original} and Quantum Approximate Optimization Algorithm (QAOA) \cite{faihi2014} are the prominent ones.

However, after a careful analysis of these methods, questions about scalability were raised with a better understanding of the phenomenon known as barren plateaus \cite{Jarrod,cerezoP,NIBP}. As a consequence, the convergence stagnation effect becomes quite critical with increasing the number of qubits. 
Recently, Ref. \cite{QCEPROC} presented a variational method based on variable circuits without the use of classical optimizers in the procedure and with a very low convergence stagnation effect. The method presented posterior good results when applied to supervised learning problems \cite{evo2023}.

The set partitioning problem is useful to modeling practical applications as, for example,  airline scheduling. This optimization problem is especially interesting for testing quantum algorithms.
The QAOA approach is generally considered more appropriate to solve combinatorial problems. The performance of QAOA was improved in \cite{qaoa_image}, a comprehensive exposition on this algorithm can be found in Ref. \cite{QIAN2023}. On the other hand, VQE is generally considered not appropriate to solve combinatorial problems, being better designed for optimizing problems related to molecular studies and spectral decomposition among other applications \cite{Kandala_2017,TILLY20221, Anton}. 

The objective of this paper is to propose and evaluate a quantum circuit evolutionary framework to solve the partitioning problem set, a well-known NP hard combinatorial optimization challenge \cite{garey97}. The framework introduces two approaches: an ansatz-free evolutionary method \cite{QCEPROC}, and a novel ansatz incorporating a pseudo-counterdiabatic term inspired by Ref. \cite{CDQO}. These methods aim to address the convergence stagnation commonly observed in variational quantum algorithms \cite{Jarrod}. 

The paper is structured as follows: the set of partitioning problem, and its formulation as a QUBO are introduced, followed by a detailed description of the proposed framework and its two approaches. Experiments were conducted under both noise-free and noisy scenarios, the performance of the proposed methods were compared with the VQE. We adopted VQE as the algorithm for comparison since we are using the same instances considered in Ref. \cite{VQE_SHORT}.
Finally, the results and their implications for scalability and convergence challenges are discussed, leading to conclusions about the potential of the proposed framework for large-scale optimization problems.

\section{The set partitioning problem}

In this section, a brief introduction to the target problem will be described. The set partitioning problem is a well-known combinatorial optimization problem which is used in several practical applications and also, is a NP-hard problem (see, e.g., Ref. \cite{garey97}) for which the optimization procedure is difficult to implement when it is scaled. It involves dividing a set of elements into mutually exclusive and exhaustive subsets such that each element belongs to exactly one subset. For example, in the airline industry, the problem is applied to crew allocation, where pilots and flight attendants must be grouped into crews to cover specific flight routes. Constraints include ensuring that each crew has the required qualifications for the aircraft, avoiding scheduling conflicts where a crew member is assigned to overlapping flights, and adhering to regulations, such as maximum flight hours. The goal is to minimize operational costs while meeting all these rules and constraints, making the set partitioning problem a valuable tool for optimizing resource management in complex systems.

\subsection{ Problem Formulation}

As input a collection of possible partitions \( P \), where each partition \( p \in P \) is associated with a non-negative weight \( w_p \), and a set of items \( I \). The output (if it exists), a selection of partitions \( p_1, p_2, \dots, p_\ell \) from the set \( P \) that covers all items in \( I \), minimizing the total weight \( w_{p_1} + w_{p_2} + \dots + w_{p_\ell} \). Each item in \( I \) must be included in exactly one of the selected partitions.

Considering each partition \( p \) in \( P \), a decision variable \( x_p \) is defined, which equals 1 if the partition \( p \) is chosen, and 0 otherwise. Additionally, \( I_p \) denotes the subset of items that belong to partition \( p \). The optimization problem is defined as
\begin{equation}
\text{Minimize} \quad \sum_{p \in P} w_p x_p
\label{eq:objective}
\end{equation}

Subject to:

\begin{equation}
\sum_{p \in P : i \in I_p} x_p = 1 \quad \forall i \in I,
\label{eq:constraint1}
\end{equation}

\begin{equation}
x_p \in \{0, 1\} \quad \forall p \in P.
\label{eq:constraint2}
\end{equation}

The constraints given by Eq. \eqref{eq:constraint1} ensures that every item will be assigned to exactly one partition \cite{VQE_SHORT}. In order to use quantum algorithms in this problem, we need to modify its expression into the well known QUBO formulation \cite{Glover2022}.

\subsection{QUBO Formulation}
Introducing penalty method based in optimization theory (see, e.g., \cite{OPT2019}), it is possible to modify the constrained problem into an unconstrained problem. As a consequence, the set partitioning problem can be formulated as a QUBO problem by introducing the appropriated constraints based in Eq.(\ref{eq:constraint1}) to the objective function (\ref{eq:objective}). The QUBO formulation to the problem can be written as
\begin{equation}
\text{minimize} \quad \sum_{p \in P} w_p x_p +  \sum_{p \in I} c_i \left( \sum_{p \in P ; i \in I_p} x_p - 1 \right)^2,
\label{eq:qubo}
\end{equation}
where $c_i$ are the penalty vector, for every item $i \in I$, $ c_i\sum_{p  \in P} w_p$. If the penalties are well defined, both problems have the same optimum. This problem can be converted into its physical counterpart, that is, the Ising Hamiltonian \cite{andrew_lucas}.

\section{Variational Quantum Eigensolver}
\begin{figure}
    \centering
   \includegraphics[width=0.5\textwidth]{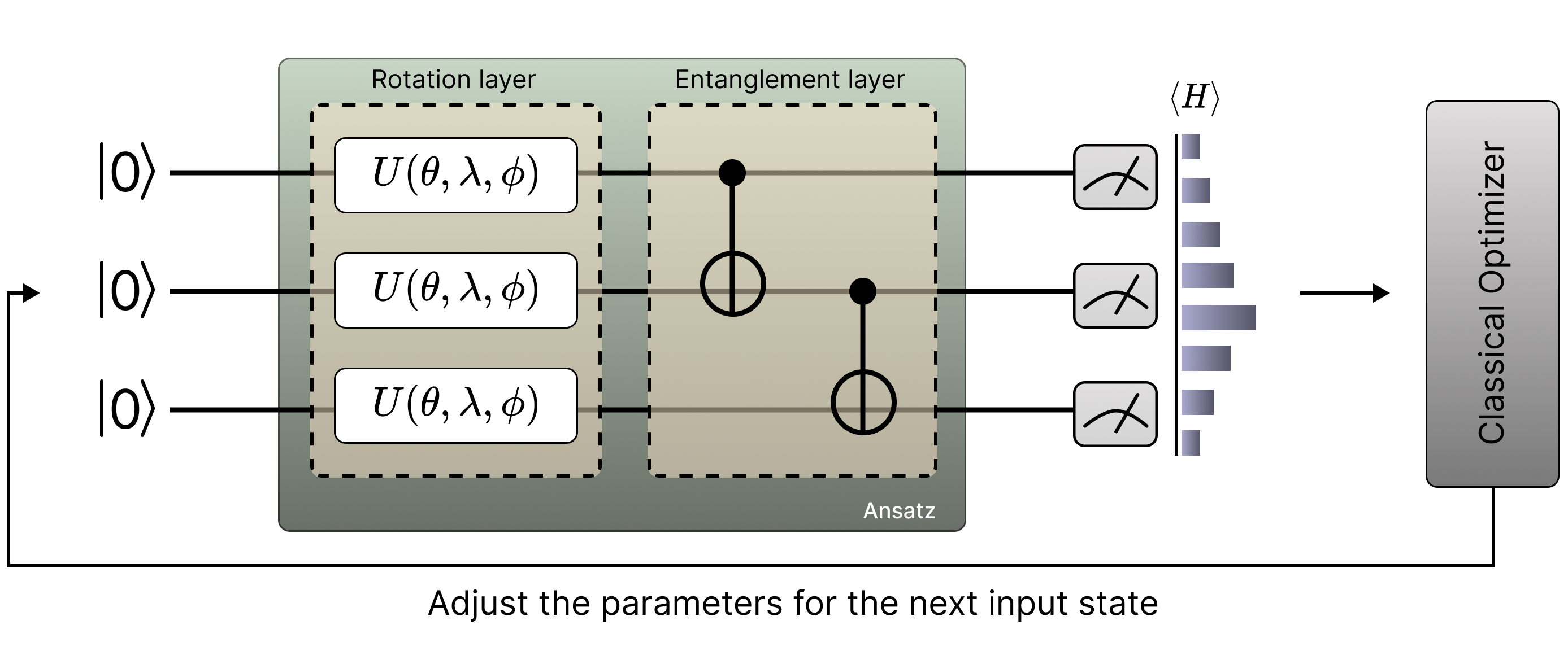}
    \caption{An illustration representing a generic fixed-ansatz quantum circuit, where the circuit structure and gate arrangement remain constant across iterations. Only the parameters of the quantum gates are updated during each iteration to optimize the circuit's performance.}
   \label{fig:1}
\end{figure}
Originally formulated for molecular optimization, the Variational Quantum Eigensolver (VQE) was first proposed by \cite{vqe_original} as an algorithm designed to find a parameterization of the quantum state, $\ket{\psi}$, that minimizes the expectation value of a given Hamiltonian. The algorithm has as target approximations of the eigenvector $\ket{\psi}$ of the Hermitian operator $H$, corresponding to its ground-state (lowest eigenvalue). This approach reduces the coherence time in order to efficiently use available  quantum resources. To achieve this minimization, the process begins with an ansatz, which is a quantum circuit made up of a series of quantum gates. On a quantum computer, these gates perform unitary operations or measurements. 

In VQE, we apply a generic parameterized unitary operator $U(\theta)$, called ansatz, to an initial state for $N$ qubits, where $\theta$ represents a set of parameters with values in the range $[-\pi, \pi]$, and the qubit register is typically initialized to the state $\ket{0}^{\otimes N}$. The VQE optimization problem can be formulated as:
\begin{equation}
 \theta=arg \min_{\theta} F(\theta),
\label{eq:thetamin}
\end{equation}
where
\begin{equation}
 F(\theta) =\bra{0}U(\theta)^{\dagger} H U(\theta) \ket{0}.
\label{eq:vqe2}
\end{equation}

This equation represents the cost function of the VQE optimization problem. The Hamiltonian, $H$, can be expressed as a weighted sum of spin operators, specifically tensor products of Pauli operators, which are observables suitable for measurement on a quantum computer. This leads to the definition of a Pauli string, $P_a \in \{I, X, Y, Z\}^{\otimes N}$, where $N$ is the number of qubits forming the wave function $\ket{\psi}$. Therefore, the Hamiltonian can be written as:

\begin{equation}
H = \sum_a w_a P_a,
\label{eq:vqe3}
\end{equation}
where $w_a$ represents a set of weights, and $P$ is the total number of Pauli strings in the Hamiltonian. Consequently, Equation [\ref{eq:vqe2}] becomes:

\begin{equation}
 \min_{\theta} \sum_a^P w_a \bra{0}U(\theta)^{\dagger} P_a U(\theta) \ket{0},
\label{eq:vqe4}
\end{equation}

Here, the hybrid nature of VQE becomes apparent: each term $E_{P_a} = \bra{0}U(\theta)^{\dagger} P_a U(\theta) \ket{0}$ represents the expectation value of the Pauli string $P_a$, which is computed on a quantum device. The summation and minimization, $E_{VQE} = \min_{\theta} \sum_a^P w_a E_{P_a}$, are then carried out using a classical optimization algorithm. A generic illustration of VQE structure is shown in Fig. \ref{fig:1}.

\subsection{Convergence stagnation}

For a large range of parametrized random circuits, expected values related to observables concentrate exponentially their gradients closely to zero \cite{Jarrod}. Such phenomenon is present even in shallow parametrized quantum circuits  \cite{cerezoP} and is called  noise-free barren plateaus. In addition to this effect, there is also the effect of noise-induced barren plateaus \cite{NIBP}. This phenomenon is one of major responsible for convergence stagnation in variational algorithms, especially in VQE. Therefore, it is expected that convergence stagnation is quite common with the increase in the number of qubits in the parametrized circuit.

\section{Quantum circuit evolutionary framework}
\begin{figure*}[!ht]
    \centering
   \includegraphics[width=0.9\textwidth]{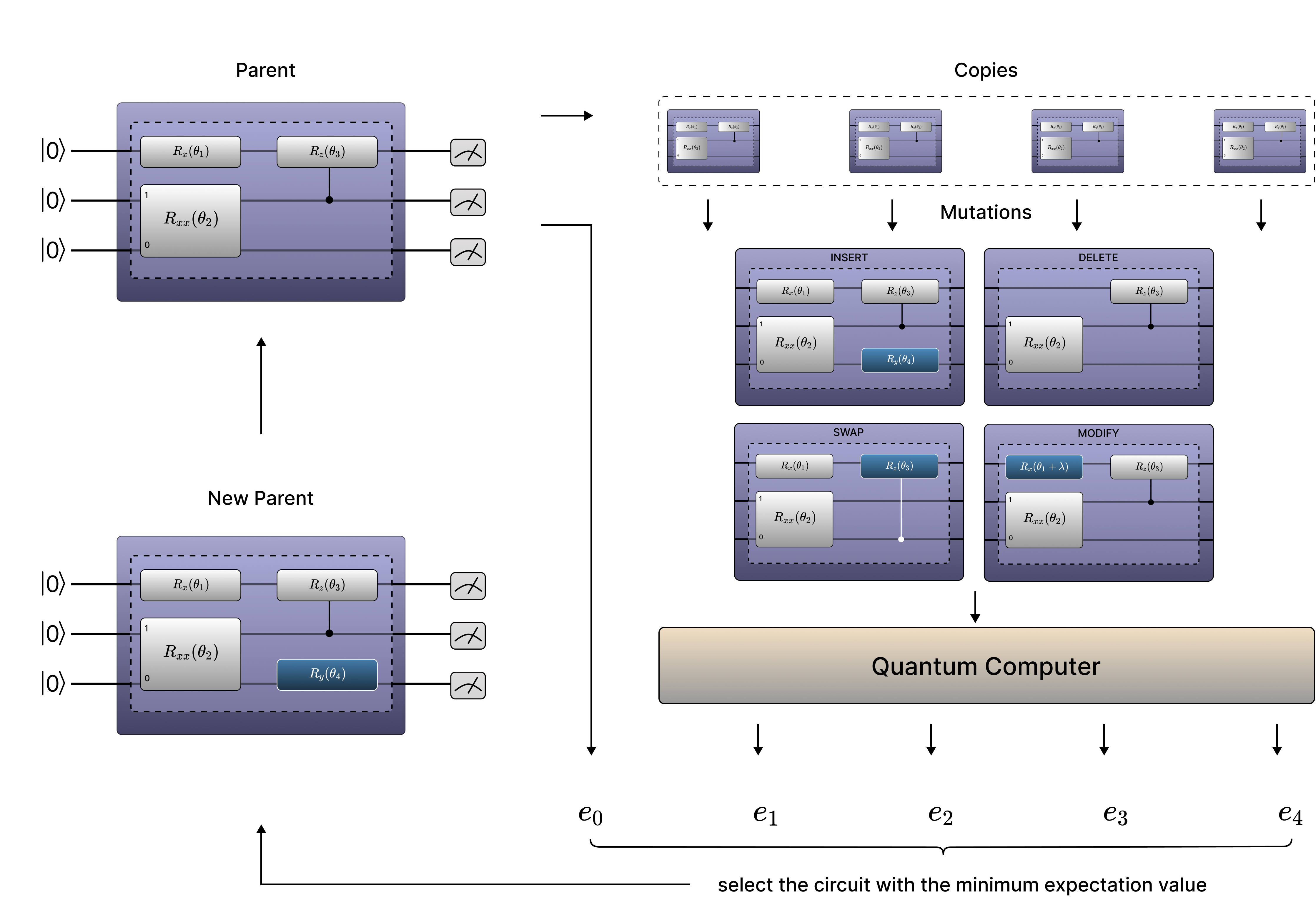}
    \caption{Workflow of the quantum circuit evolutionary process. (Left) The algorithm begins with an initialized parent circuit and its associated expectation value. Offspring circuits are generated by applying mutations to copies of the parent. These circuits are evaluated on the quantum machine, and the circuit with the lowest expectation value—whether an offspring or the original parent—becomes the new parent for the next generation. The process iterates until a convergence criterion is met. (Right) Details of the mutation strategy: INSERT (add a random gate), DELETE (remove a random gate), SWAP (replace a gate with a random new gate), and MODIFY (adjust the parameter of a random gate).}
   \label{fig:2}
\end{figure*}

The VQE mechanism works by maintaining a fixed circuit structure with varying parameters. The reference \cite{QCEPROC} introduces a different way to search the optimum. In their approach, all the circuit structure (or topology) varies in the optimum search.  in contra-position to the VQE, such method can be ansatz-free and performs updates in all circuit configuration. In other words, we have now a circuit cost function $F$ given by 
\begin{equation}
 F(\mathbf{U})=\langle\psi_0|\mathbf{U}^{\dagger}{H}\mathbf{U}|\psi_0\rangle,
\label{eqqce}
\end{equation}
where $\mathbf{U}$ is a circuit, ${H}$ is the Hamiltonian of problem and $|\psi_0\rangle$ is an arbitrary initial state. In the reference that introduced the topic, the choice of the initial state was arbitrary. In the case of this work we will evaluate two different strategies. The first one is in the same sense that Ref. \cite{QCEPROC} using an evolutionary approach starting from the state $\ket{0}^{\otimes N}$. In the second, we address our contribution where we consider an approach inspired on adiabatic evolution taking into account the corrections provided by the adiabatic gauge potential \cite{JAPA}. We call both protocols Quantum Circuit Evolution (QCE) (in the same way that \cite{evo2023}) with an ansatz-free (AF-QCE) and an ansatz with pseudo-counterdiabatic evolutionary term (APCD-QCE).
\subsection{Ansatz-free evolutionary approach}

The Ansatz-free evolutionary approach used in this work closely follows the proposal of Ref. \cite{QCEPROC}, with some modifications that are detailed in this section. This approach leverages the flexibility of quantum circuits, allowing them to adapt dynamically without relying on a pre-defined ansatz. Instead of starting with a static circuit configuration, this method begins with a minimal circuit, often consisting of just a single gate, and evolves iteratively through a process of mutations.

The initial circuit is composed of a single gate, which can be one of [$R_x(\theta)$, $R_y(\theta)$, $R_z(\theta)$], where $\theta \in [0, 2\pi]$. The gate type, the angle $\theta$, and the qubit to which the gate is applied are all chosen randomly. Following this, mutation operations play a crucial role in the evolutionary process.

The gates used in this algorithm are divided into two main categories: single-qubit gates and two-qubit gates. The single-qubit gates include rotations around the x, y, and z axes, represented by $R_x$, $R_y$, and $R_z$, respectively. The two-qubit gates group includes controlled rotations and direct interactions between two qubits, such as $R_{xx}$, $R_{yy}$, and $R_{zz}$, as well as controlled rotation gates $CR_x$, $CR_y$, and $CR_z$. Possible mutation operations, each with an assigned probability weight, are as follows \cite{qce}:

\begin{itemize}
\item \textbf{INSERT} (0.25): Insert a new gate $g$ with a random $	\theta$ at a random position within the circuit.
\item \textbf{DELETE} (0.25): Delete an existing gate from a randomly selected position.
\item \textbf{SWAP} (0.25): Combine a delete and an insert operation at the same randomly chosen position.
\item \textbf{MODIFY} (0.25): Modify the parameter of a randomly chosen gate by adjusting $	\theta$ according to $\theta \rightarrow \theta + \epsilon$, where $\epsilon \sim N(0, 0.1)$.
\end{itemize}

Figure \ref{fig:2} illustrates the optimization process, for each generation a population of four circuits is created by making copies of a parent circuit and applying random mutations to each copy. The mutations are designed to thoroughly explore the search space of possible quantum circuits, ensuring that no significant restrictions are placed on the evolution of the circuit's structure. This Ansatz-free approach, starting with a single gate, allows the algorithm to maintain greater adaptability, enabling the circuit to evolve to better suit the specific optimization problem at hand.

The parent circuit and its four mutated copies are measured on a quantum computer. From these, the best-performing circuit is selected to serve as the new parent for the next generation. This ensures that the evolution process is monotonic, with the value of the cost function either decreasing or staying the same over successive generations. The optimization steps are repeated until convergence is achieved or a maximum number of iterations is reached.

This strategy of continuous adaptation and selection ensures that the quantum circuit evolves toward an optimal configuration, offering significant advantages over fixed-ansatz methods by adapting the circuit architecture to the problem dynamically.

\subsection{Ansatz using trotterization with pseudo-counterdiabatic evolutionary terms}
Let us now consider an ansatz inspired on the physics of the problem to try improve the convergence of the method. To make it, we define an ansatz related to the Hamiltonian of the problem taking the  time Hamiltonian  given by 
\begin{equation}
    H_{ad}(t)=[1-\lambda(t)]H_{0}+\lambda(t)H_{I}.
\end{equation}

Here, the scheduling function is $\lambda(t)\in[0,1]$, $H_{I}$ and $H_0$ are, respectively, the Hamiltonian of the problem and a given initial Hamiltonian from which is easy to prepare the ground-state vector. Besides, the adiabatic theorem (see, e.g.,\cite{faihi2014}) ensures that the system is preserved in the instantaneous ground-state if the  time evolution is sufficiently slow.  

\begin{figure}
    \centering
   \includegraphics[width=0.5\textwidth]{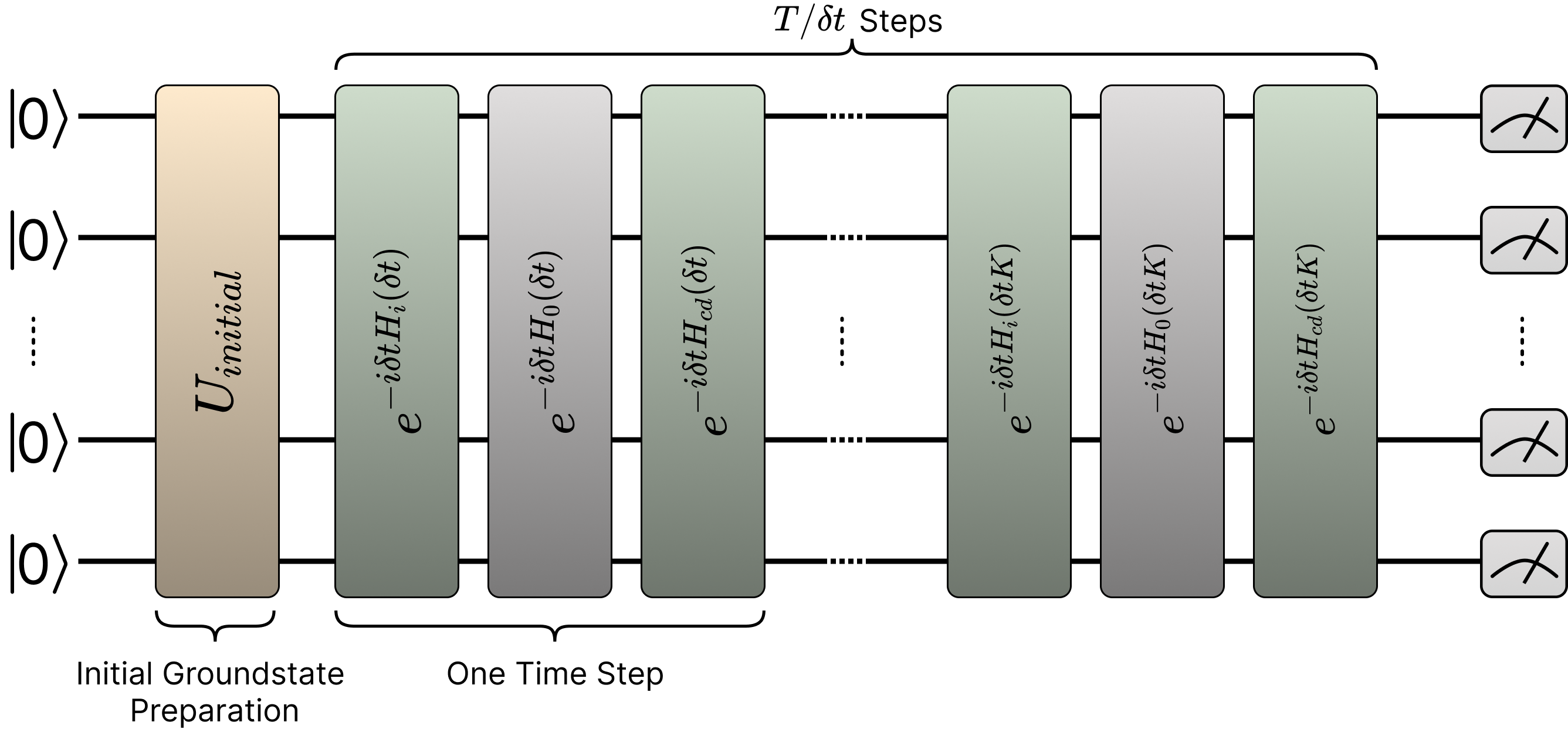}
    \caption{Illustration the digitalized counterdiabatic protocol introduced in \cite{CDQO}.}
   \label{fig:3}
\end{figure}

Also, in order to create a digitalized circuit, it is necessary make the trotterization described in reference \cite{Suzuki}. Such a procedure is the basic tool to create the digitalized QAOA \cite{faihi2014}. The introduction of counterdiabatic terms is an option to speed up adiabatic evolutions \cite{CDQO}. The digitized counterdiabatic quantum optimization protocol is illustrated in Fig. \ref{fig:3}. Furthermore, a variation of this approach based on an impulse regime (characterized by a classical optimization) was presented in \cite{cadavid2023}.

In reference \cite{JAPA}, algebraic expressions for the adiabatic gauge potential (AGP) were introduced. As a consequence, this term describes non-adiabatic transitions and, if such term is added to a time evolution Hamiltonian with proper conditions the last may obtain good results even in the presence of regimes where we have short time evolution. 

This type of tool is particularly important for applications in the current NISQ devices since they have a short coherence time. Counterdiabatic terms are extracted from variational minimization of action which is given in terms  of the Hilbert-Schmidt operator. With appropriate calculations it is possible to obtain a first order approximate solution  with good practical results \cite{JAPA,CDQO,cadavid2023}.

In order to improve the QCE convergence, we introduce the pseudo-counterdiabatic evolutionary term to define the ansatz. The last is defined as $U=U_{ad}U_{pcd}$ inspired on the total Hamiltonian $H_{T}$ which is given by
\begin{equation}
    H_{T}(\lambda)=H_{ad}(\lambda)+H_{pcd}(\lambda),
\end{equation}
where $H_{ad}$ is inspired in the adiabatic time evolution Hamiltonian and  $H_{pcd}$ is the pseudo-counterdiabatic evolutionary term. The operator  $U_{ad}$ will be fixed and define as one trotter step (or one layer if we thinking similarly to QAOA procedure) from the digitalized counterdiabatic protocol. On the other hand, the operator $U_{pcd}$ will be evolutionary in preserving the structure of QCE. In other words, we have
\begin{equation}
 U_{ad}(\beta,\delta)=e^{-i\beta H_{I}} e^{-i\delta H_{0}}=U_{I}(\beta)U_{0}(\delta)
\end{equation}
and, in order to define constraints on how its evolution is setting, we define 
\begin{equation}
    U_{pcd}= e^{-i\theta_{i}H_{pcd}},
\end{equation}
with $\theta_i$ given randomly and $H_{pcd}$ randomly generated from the basis $\{Y,I\otimes X, I\otimes Y,I\otimes Z,X\otimes X, Y\otimes Y, Z\otimes Z\}$. Therefore, $U_{pcd}$ is obtained from the set of gates $\{R_y, CR_x,CR_y,CR_z,R_{xx}, R_{yy}, R_{zz}\}$ using the evolutionary approach. 

\begin{figure*}[!ht]
    \centering
   \includegraphics[width=0.8\textwidth]{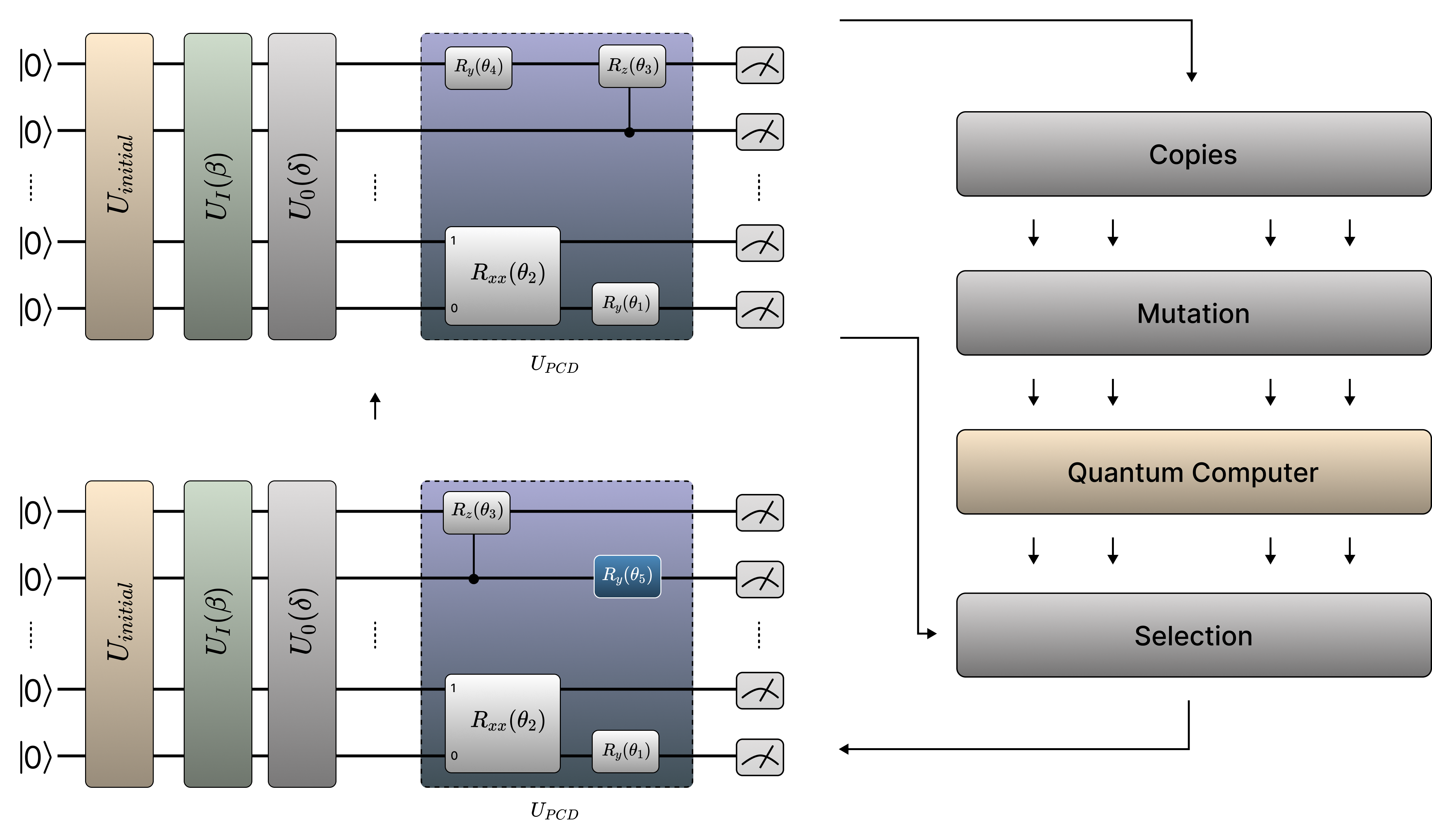}
    \caption{Scheme of the workflow for the APCD-QCE configuration.}
   \label{fig:4}
\end{figure*}

The procedure is illustrated in Fig. \ref{fig:4}. Note that the choice of a trotter step is based on the purpose of not obtaining a very deep circuit. However, the aforementioned scheme can be adapted for trotter step repetitions in an analogous way to the protocol described in Fig. \ref{fig:3}.

\textbf{Remark:} Here we reinforce that the physics of the problem is only inspiring the construction of our ansatz, which is a guess by definition, and we do not use any physical parameters in the construction.

\section{Simulator experiments}

Computational simulations were conducted on the quantum simulator KUATOMU at SENAi CIMATEC`s High-Performance Computing Center (HPC). The machine consists of 192 processing cores (CPU) with 384 threads, 3TB of RAM memory and 4 NVidia V100 32GB GPU accelerator cards. We use the open-source library Qiskit in our implementations.

The experiments were divided into two categories: noise-free simulations and noise simulations, which we will explain with details in the following subsections. In both cases, seven runs with 10,000 iterations (or generations for QCE) each were conducted for each instance of the set partitioning problem. In Table \ref{tab:my-table_0}, each instance name corresponds to the number of qubits that was designed to evaluate the performance of quantum algorithms. The VQE, AF-QCE and APCD-QCE algorithms were applied to solve the problem.

The VQE's ansatz used was a two-local circuit with linear entanglement and a fixed number of repetitions set to 2. The number of measurement shots was set at 1024 and the optimizer used for the variational parameters was COBYLA \cite{COBYLA}, with a convergence tolerance set to $10^{-6}$. This setup was chosen to align as closely as possible with the approach in Ref. \cite{VQE_SHORT}, which is also addressed to solve the same problem.

The AF-QCE was set considering all the specifications described in the previous section. On the other hand, for the APCD-QCE, we use the initial preparation $|+\rangle^{\otimes n}$ with the unitary operators running considering $\beta=0$ and $\delta=0.5$ and $U_{0}(\delta) = \bigotimes_{i=0}^{n}R_{x_i}(2\delta)$.

To evaluate the algorithms, we used a metric based on the median of the expected value normalized by a reference value calculated using the classical Gurobi solver. This metric, which ranges from 0 to 1, provides an intuitive measure of how closely the quantum algorithms approach the optimal solution, with values closer to 1 indicating better performance. This efficiency measure is called approximation ratio $\cal{R}$ and defined by
$\cal{R}=\text{Reference value}/\text{Minimum expectation value}$. Since the small value of reference  regarding all instances is $20$, there is no possibility for $\text{Reference value}$ equals zero resulting in an indetermination when the minimum is achieved.

\renewcommand{\arraystretch}{1.2} 
\newcommand\myVSpace[1][10pt]{\rule[\normalbaselineskip]{0pt}{#1}}

\begin{table}[]
\caption{The VQE, AF-QCE, and APD-QCE results on the benchmark set partitioning instances of Svensson et al. \cite{database}for the noise-free scenario.}
\label{tab:my-table_0}
\scalebox{0.9}{}{%
\begin{tabular}{|llll|}
\hline
\myVSpace[3pt]Instances \hspace{20pt}  & \multicolumn{3}{l|}{Approximation ratio (${\cal{R}}$)}  \\  \cline{2-4} 
\myVSpace[3pt]          & VQE \hspace{16pt}      & AF-QCE  \hspace{16pt}      & APCD-QCE \hspace{16pt} \\ \cline{1-4} 
\myVSpace[3pt]6.1       & \textbf{1.00}        & \textbf{1.00}             & \textbf{1.00}         \\ 
6.2       & 0.39        & \textbf{1.00}        & \textbf{1.00}         \\ 
6.3       & 0.86        & 0.92                 & \textbf{1.00}         \\
\myVSpace[3pt]8.1       & 0.05        & \textbf{1.00}        & \textbf{1.00}         \\ 
8.2       & 0.33        & 0.33                 & \textbf{1.00}         \\ 
8.3       & 0.87        & 0.87                 & \textbf{1.00}         \\ 
8.4       & 0.45        & \textbf{1.00}        & \textbf{1.00}         \\ 
\myVSpace[3pt]10.1      & 0.02        & \textbf{1.00}        & \textbf{1.00}         \\ 
10.2      & 0.01        & 0.19                 & \textbf{1.00}         \\ 
10.3      & 0.02        & \textbf{1.00}        & \textbf{1.00}         \\ 
10.4      & 0.01        & 0.74                 & 0.98                  \\ 
10.5      & 0.02        & 0.97                 & \textbf{1.00}         \\ 
\myVSpace[3pt]12.1      & 0.03        & 0.09                 & \textbf{1.00}         \\ 
12.2      & 0.01        & \textbf{1.00}        & \textbf{1.00}         \\ 
12.3      & 0.01        & 0.46                 & 0.99                  \\ 
12.4      & 0.01        & 0.70                 & \textbf{1.00}         \\ 
12.5      & 0.01        & 0.44                 & \textbf{1.00}         \\ 
12.6      & 0.01        & 0.43                 & 0.65                  \\ 
\myVSpace[3pt]14.1      & 0.01        & 0.07                 & \textbf{1.00}         \\ 
14.2      & 0.01        & 0.90                 & \textbf{1.00}         \\ 
14.3      & 0.02        & 0.76                 & \textbf{1.00}         \\ 
14.4      & 0.00        & 0.47                 & \textbf{1.00}         \\ 
14.5      & 0.00        & 0.59                 & 0.95                  \\ 
14.6      & 0.01        & \textbf{1.00}        & 0.96                   \\ 
14.7      & 0.01        & 0.77                 & 0.95                   \\ 
\myVSpace[3pt]20.1      & 0.00        & 0.06                 & 0.06                   \\ 
20.2      & 0.00        & 0.44                 & \textbf{1.00}                   \\ 
20.3      & 0.00        & 0.42                 & 0.90                   \\ 
20.4      & 0.00        & 0.20                 & \textbf{1.00}                   \\ 
20.5      & 0.00        & \textbf{1.00}        & 0.31                   \\ 
20.6      & 0.00        & 0.14                 & 0.47                   \\ 
20.7      & 0.00        & 0.45                 & 0.46                   \\ 
20.8      & 0.00        & 0.37                 & 0.61                   \\ 
20.9      & 0.00        & 0.71                 & 0.90                   \\ 
20.10     & 0.00        & 0.65                 & 0.74                   \\ \cline{1-4} 
\end{tabular}%
}
\end{table}


\subsection{Noise-free simulator experiments}

In the case of noise-free simulations, all 35 instances of the set partitioning problem were evaluated, as shown in Table \ref{tab:my-table_0}. These simulations provided a baseline for analyzing the idealized performance of the algorithms without the influence of quantum noise.
In this scenario, in the majority cases, the AF-QCE and APCD-QCE outperform the VQE algorithm with respect to the chosen metric. When we compare the two approaches with QCE, we see that APCD-QCE has better convergence in most of the instances studied. 

\begin{figure}[!ht]
   \includegraphics[width=0.45\textwidth]{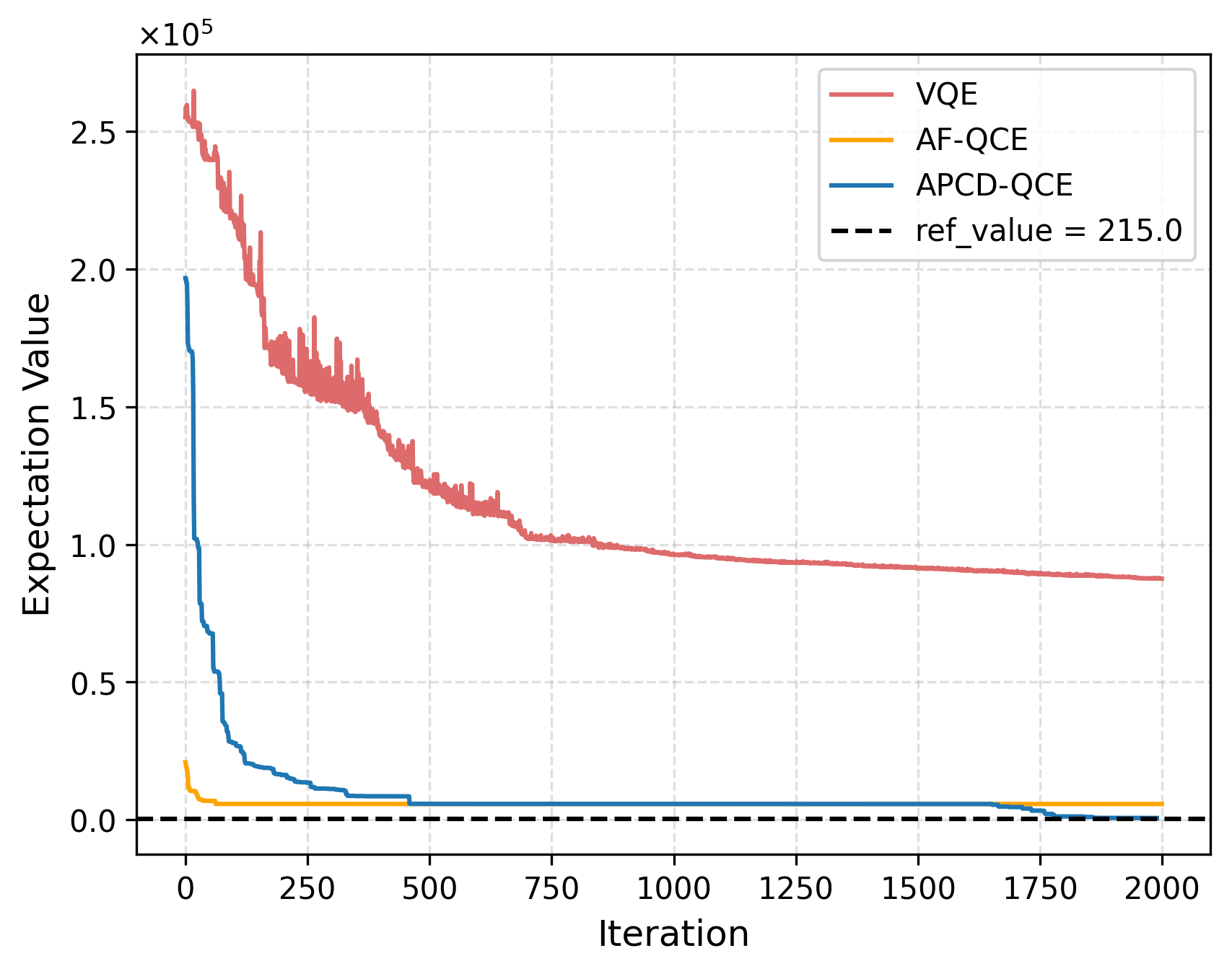}
    \caption{Convergence performance for 14 qubits (a zoomed-in view of the interaction range was applied for a more detailed observation). In this run, APCD-QCE presents better performance compared to AF-QCE and VQE.}
   \label{fig:convergence}
\end{figure}

The performance of both the AF-QCE and APCD-QCE approaches shows diminished convergence as the number of qubits increases. In particular, we see very poor performance for the instances $20.1$, $20.6$ and $20.7$. In these cases, convergence stagnation is evident for both approaches. In Fig. \ref{fig:convergence}, the stagnation of convergence for VQE in a specific run is illustrated. For this same run, the AF-QCE and APCD-QCE avoid the stagnation.
\subsection{Noise simulator experiments}

In real case scenarios, all the proposed kind of algorithm will be challenged in presence of noise and errors. As it is a sensitive physical system, errors appear in the preparation of states as well as in the carrying out of measurements. Of course, these types of errors will have a noticeable effect on methods using multiple measures of the expected value. Another type of error is related to
the interaction of the physical qubits with the surrounding environment. This type of error was not considered in our analysis either.  For this scenario, we will evaluate errors and noise for qubits $1$ and $6$. Therefore, it was considered a depolarizing error channel (which is described by errors in Pauli product \cite{noisygates}) with an error $\epsilon=0.01$  for the elected gate $CY$.

\begin{figure}[!ht]
   \includegraphics[width=0.36\textwidth]{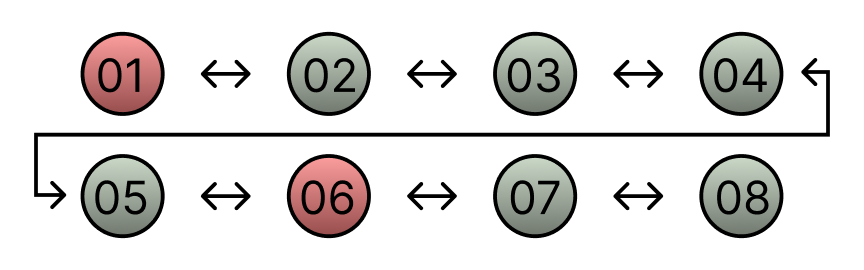}
    \caption{Coupling map setup with the red ones representing the problematic qubits.}
   \label{fig:5}
\end{figure}

Figure \ref{fig:5} illustrates the linear qubit coupling map generated our noise simulations. In this configuration, the qubits are arranged in a sequential layout where each qubit is connected only to its immediate neighbors. Specifically, qubit 1 is coupled with qubit 2, qubit 2 with qubit 3, and so on. This linear connectivity reflects a simple topology that supports two-qubit gates between adjacent qubits.
\renewcommand{\arraystretch}{1.2} 
\begin{table}[]
\caption{The VQE, AF-QCE, and APD-QCE results on the benchmark set partitioning instances of Svensson et al. \cite{database} for the noise scenario.}
\label{tab:my-table_1}

\scalebox{0.9}{}{%
\begin{tabular}{|llll|}
\hline
\myVSpace[3pt]Instances \hspace{20pt}  & \multicolumn{3}{l|}{Approximation ratio ($\cal{R}$)}  \\  \cline{2-4} 
\myVSpace[3pt]          & VQE \hspace{16pt}      & AF-QCE  \hspace{16pt}      & APCD-QCE \hspace{16pt} \\ \cline{1-4} 
\myVSpace[3pt]6.1       & 0.06        & \textbf{1.00}             & \textbf{1.00}         \\ 
6.2       & 0.04        & \textbf{1.00}        & \textbf{1.00}         \\ 
6.3       & 0.73        & 0.92                 & 0.92         \\
\myVSpace[3pt]8.1       & 0.03        & 0.10        & \textbf{1.00}         \\ 
8.2       & 0.01        & \textbf{1.00}                 & 0.33         \\ 
8.3       & 0.02        & 0.87                 & 0.90         \\ 
8.4       & 0.02        & 0.45        & 0.17         \\ \cline{1-4} 
\end{tabular}%
}
\end{table}

 As we have the intention of evaluate the effect of defined noise on the algorithms, we decided to proceed with the analysis for the cases of six and eight partitions only. Table \ref{tab:my-table_1} shows the results in this context, where we see  a downgrade in the performance of VQE and QCE in terms of our convergence metric. Overall, as illustrated in Table \ref{tab:my-table_1}, both QCE frameworks have a good performance for the majority of instances with comparable results.

\section{Discussion}
The results presented in Table \ref{tab:my-table_0} highlight significant differences in the performance of the VQE, AF-QCE, and APCD-QCE algorithms under the noise-free scenario. These findings provide critical insights into the relative effectiveness of each approach and the challenges faced as the number of qubits increases.

In the idealized noise-free scenario, APCD-QCE consistently demonstrated the best performance, achieving the approximate ratio closest to 1 in 25 out of 35 evaluated instances. In comparison, AF-QCE outperformed VQE in most instances, except for a few cases where both struggled to converge. The VQE, although effective in some smaller instances (e.g., 6.1), showed significantly degraded performance as the number of qubits increased, with consistently low results for instances of 10 qubits or more. 

This poor performance can be attributed to the barren plateaus effect, making it challenging for the optimizer to find a meaningful direction for improvement. This phenomenon is explicit in Figure \ref{fig:convergence}, which illustrates the convergence stagnation for VQE as the cost function gradients approach zero in higher-dimensional instances. 

As discussed in the reference \cite{Otfree}, even the use of gradient-free optimizers does not allow us to efficiently mitigate these intrinsic effects of parameterized circuits. APCD-QCE outperformed AF-QCE in most instances, evidencing its superior ability to avoid convergence stagnation. For example, in instances such as 10.2 and 12.1, AF-QCE showed relatively low approximation values (0.19 and 0.09, respectively), while APCD-QCE reached the maximum value of 1.00. 

This advantage can be attributed to the initial state preparation $|+\rangle^{\otimes n}$ and specific parameter settings ($\beta = 0$, $\delta = 0.5$) used in APCD-QCE, which seem to facilitate a more effective exploration of the solution in  Hilbert's space. In fact, with such settings, the expression of the Hamiltonian of problem does not has influence in the ansatz. 

Therefore, the initial Hamiltonian $H_0$  with easy ground state preparation has a major influence in such configuration. The performance of all algorithms degraded for larger instances, particularly those with 20 qubits. For instances such as 20.1, 20.6, and 20.7, both AF-QCE and APCD-QCE struggled to converge, with approximation values close to zero for 20.1 and below 0.5 for the other two. 

These results indicate that even in a noise-free setting, the algorithms face significant scalability challenges, likely due to the increasing complexity of the problem as the number of qubits increases to 20. On the other hand, introducing noise led to reduced performance, particularly in small instances, given clues that noise, just as in parameterized circuits, can affect the convergence of techniques based in variable circuits with fixed parameters.

A critical issue in the proposed approach is the number of measures, since the procedure uses an evolutionary approach and, consequently, more circuits than usual VQA's. In the approach chosen for this paper, we use four circuits as the initial population, resulting in the number of shots multiplied by this value which is greater than used for VQE. However, the convergence stagnation present in the VQE is intrinsic to the parametrization of the chosen ansatz being not critically affected by the number of shots. As a consequence, increasing the number of measurements does not significantly change the convergence performance of this algorithm. Furthermore, the convergence difficulties in this category of algorithms are closely related to the inability of the classical optimizer to avoid local minima. With respect to QCE, for the $20-$ qubits instances, the number of shots remained unchanged in relation to the other instances, being only $1024$ shots per circuit, for a set of more than a million possible amplitudes. 

Therefore, despite some difficulties in convergence, we considered the presented results satisfactory. Another thing to be considered in this discussion, VQE had its performance improved for some instances in the reference \cite{VQE_SHORT}. In that work,  it was used  reductions and smart penalties in order to reduce the number of variables. This approach is also very useful for improving the efficiency of some classic algorithms. In our approach, no kind of relaxation was performed on the instances showing the results of each methodology for the dimension of the problem studied.

\section{Conclusions}
In this paper,  a framework based on quantum circuit evolution for solving instances of the set partitioning problem was presented. The proposed framework uses quantum circuit evolutionary topology based in ansatz-free evolutionary technique and, inspired on counterdiabatic protocols, an additional strategy using an ansatz with a evolutionary pseudo-counterdiabatic term. Both strategies were compared with the VQE achieving better performance in relation to the last. 

However, in all instances, it was observed that in some of the runs, the problem of stagnation occurs. It seems that the mutation strategy can lead to a convergence to a circuit configuration where, within the Hilbert's space, neighboring circuits do not lead to an improvement in the cost function.  
Escaping this local minimum stagnation can be challenging, as mutations occur incrementally, one at a time. This behavior is similar to the barren plateaus but less common, indicating that this procedure is almost immune to convergence stagnation effects with a sufficiently large number of generations. 

Furthermore, APCD-QCE demonstrated greater resilience in avoiding stagnation, as illustrated in Fig. \ref{fig:convergence}. In cases where VQE and AF-QCE became stuck in suboptimal values, APCD-QCE managed to achieve the optimal solution, underscoring its effectiveness in mitigating this problem.

\section{Acknowledgments}
This work has been partially supported by QuIIN - EMBRAPII CIMATEC Competence Center in Quantum Technologies, with financial resources from the PPI IoT/Manufatura 4.0 of the MCTI grant number 053/2023, signed with EMBRAPII. Also, this work was partially financed by CNPq (Grant Numbers: 305096/2022-2, Marcelo A. Moret). We also thank Anton Simen Albino for discussions on AF-QCE.

\textbf{Code availability statement:} The code used in this work is available at: \href{https://github.com/brunooziel/quantum-circuit-evolutionary-framework-applied-on-set-partitioning-problem}{https://github.com/brunooziel/quantum-circuit-evolutionary-framework-applied-on-set-partitioning-problem}.

\bibliography{sample}
\end{document}